# Compiling molecular ultrastructure into neural dynamics

**Konrad P. Kording, Anton Arkhipov, Davy Deng, Sean Escola, Seth G.N. Grant, Gal Haspel, Michał Januszewski, Narayanan Kasthuri, Nina Khera, Richie E. Kohman, Grace Lindsay, Jeantine Lunshof, Adam Marblestone, David A. Markowitz, Jordan Matelsky, Brett Mensh, Patrick Mineault, Andrew Payne, Joanne Peng, Xaq Pitkow, Philip Shiu, Gregor Schuhknecht, Sven Truckenbrodt, Joshua T. Vogelstein, Edward S. Boyden**

**Abstract** High-resolution brain imaging can now capture not just synapse locations but their molecular composition, with the cost of such mapping falling exponentially. Yet such ultrastructural data has so far told us little about local neuronal physiology – specifically, the parameters (e.g., synaptic efficacies, local conductances) that govern neural dynamics. We propose to translate molecularly annotated ultrastructure into physiology, introducing the concept of an ultrastructure-to-dynamics compiler: a learned mapping from molecularly annotated ultrastructure to simulator-ready, uncertainty-aware physiological parameters. The requirement is paired training data, with jointly acquired ultrastructure from imaging, and dynamical responses to perturbations from physiological experiments. With this data we can train models that predict local physiology directly from structure. Such a compiler would support biophysical simulations by turning anatomical maps into models of circuit dynamics, shifting structure-to-function from a descriptive program to a predictive one and opening routes to understanding neural computation and forecasting intervention effects.

# Introduction

A compiler translates abstractions of a computer program in a source language into executable mechanisms in a target language[1]. Neuroscience now faces an analogous problem: we can image the molecular and ultrastructural components of neural tissue in extreme detail, visualizing many molecules and even interactions, but we cannot yet convert what we see into mechanistic models that reliably reproduce the dynamics we measure. We call this desired translation system an "ultrastructure-to-dynamics compiler" (or UDC for short; Box 1), which maps molecularly organized ultrastructure into distributions over simulator-ready parameters with calibrated uncertainty. Here, ultrastructure does not mean geometry alone. The relevant source language includes not only connectivity and morphology, but also molecular identity and molecular organization across scales, including subsynaptic and nanoscale arrangements. Neuroscience has historically inferred mechanisms from activity by fitting hidden parameters to match observations. Much of neuroscience has therefore depended on an underdetermined[2] inverse problem: many distinct parameter settings can explain the same observed activity, and more data alone does not resolve that ambiguity.

> **Box 1: Ultrastructure-to-dynamics (UDC) compiler**
> **Input:** datasets of molecularly annotated ultrastructure (wiring, morphology, molecular markers) plus context metadata.
> **Output:** simulator-ready distributions over effective synaptic and compartment parameters (conductances, kinetics, plasticity, noise), with calibrated uncertainty.
> **Built from:** paired datasets that link molecularly annotated ultrastructure to measured local dynamics under controlled perturbations (e.g. through patchclamping, optical physiology), trained with machine learning.

The inverse problem is quite hard relative to the much simpler *forward* problem of simulating a circuit[3]. The status quo is therefore changing as the causal machinery of the nervous system becomes observable at scale. Our proposal is not to abandon inverse inference that drives modern neuroscience, but to constrain and steer it by translating observed structure, molecular identity, and molecular organization into the parameters that drive dynamics.

An ultrastructure-to-dynamics compiler uses physical constraints to collapse the space of plausible causal models (see Fig. 1). If we can infer distributions over effective synapse and compartment parameters from imaging alone, those local models can be translated into forward simulations that scale from cells to circuits. Here, we outline the ultrastructure-to-dynamics compiler and the research program needed to build it. We propose to learn this translation from paired datasets linking molecular ultrastructure to measured local dynamics, allowing images to be compiled into models that predict responses to drugs, genetic perturbations, and stimulation. Done well, this approach will also reveal which structural and molecular details matter for which predictions, enabling principled abstraction rather than ad hoc simplification (see Box 2 for targets and metrics).

Feasibility is supported by adjacent successes that show both sides of the ultrastructure-to-dynamics bridge. On the structural side, ultrastructural features already carry measurable functional signals: synapse size and cell-type context predict a substantial fraction of postsynaptic potential (PSP) variance in some classes[4], and neurotransmitter identity can be inferred from EM[5]. At the same time, similar-looking synapses can behave differently, and dendritic context and intrinsic excitability shape how local events appear at the soma. This is why the compiler must output distributions over effective parameters[6], including stochasticity and uncertainty.

On the modeling side, mechanistic neuron models can be fit automatically from morphology and electrophysiology at scale[7–11], and transcriptomic identity predicts intrinsic electrophysiological

> **Box 2: Targets and metrics**
> Success should be scored with a set of preregistered, out-of-distribution benchmarks at three nested scales.
> **Synapse level.** From molecularly annotated ultrastructure, predict the distribution of PSP amplitudes and kinetics, including short-term plasticity and stochastic release. Test on held-out pharmacology and release perturbations, scoring predictive error and uncertainty calibration.
> **Cell level.** From morphology and molecular context, predict intrinsic electrophysiology determining excitability and integration (e.g., f–I curve, subthreshold impedance, adaptation) and their changes under held-out channel modulators. Score accuracy and calibration on held-out cell types and perturbations.
> **Circuit level.** Translate compiled synapse and neuronal compartment models into circuit and whole-system simulations, and predict changes in population activity under held-out interventions (for example optogenetic or sensory perturbations), scoring match to observed changes in firing rate, synchrony, and other measurements.
> Benchmarks should be pre-registered and agreed upon by the community before results are announced.

phenotype and places informative priors on local connectivity patterns, even within broad interneuron classes[12–14].

Even "simple" cases show why ultrastructure-to-dynamics is hard. Take an identified glutamatergic synapse onto a pyramidal neuron. The protein composition, organisation and turnover of excitatory synapses vary widely[15]. The translation from local structure to a somatic excitatory postsynaptic potential (EPSP) depends heavily on compartmental context: channel densities and kinetics set excitability, while dendritic geometry determines nonlinear integration and all of these are highly heterogeneous[16]. Furthermore, in intact tissue, effective parameters are dynamically shifted by neuromodulatory state[17]

Crucially, this is why purely structural connectomics is insufficient: functional prediction depends not just on connectivity and morphology, but on molecular identity and molecular organization, including nanoscale arrangements that shape transmission, integration, and state dependence[19,20]. In addition to receptor subtype and signaling machinery, the supramolecular organization of postsynaptic proteins, receptor nanodomains, and pre/post alignment can shape effective synaptic strength and kinetics[21,22,20]. A static snapshot will not reveal the instantaneous neuromodulatory state, but molecular readouts can provide priors on how parameters shift across states (for example via receptor subtypes, transporters, and synthesis machinery). The compiler therefore does not ignore volume transmission. It outputs conditional parameter distributions, learned from paired datasets in which neuromodulatory conditions are experimentally varied, so that state dependence is represented as uncertainty or as explicit covariate sensitivity rather than assumed away. Validation is then direct: predict how response distributions shift under established

> **Box 3: Medical relevance**
> CNS drug development can confirm that a molecular compound affects its target but cannot reliably predict how the resulting cascade of signals reshapes circuit dynamics; which cell types are affected, how excitation–inhibition balance shifts, or why responses vary across brain regions and states. Thus clinical trials often fail at a late stage when a drug's effect doesn't generalize across the varying states and contexts that a human patient might present, and side effects are also commonly detected, at such a late stage of investigation, often wasting a decade and billions[18]. An ultrastructure-to-dynamics compiler would close this gap: given molecularly annotated tissue, model the circuit-level consequences of a receptor-level intervention and test those predictions. Then efficacy could be more robustly predicted, and perhaps even side effects anticipated - replacing late stage risk with earlier insight. The same logic applies to any intervention where the question is not whether a molecule is present, but how changing it alters the dynamics of the circuit it sits in.

pharmacological and other perturbations. If we can compile measured components into dynamics, we can forecast how interventions alter neural dynamics and propagate through circuits, a prerequisite for predicting therapeutic effects. It would enable executable models of circuits and, eventually, whole brains (see [23–27]). Molecular imaging is detailed and affordable enough to support a bottom-up workflow: observe the components, infer the parameters of mechanistic models, and predict circuit responses at larger scales.

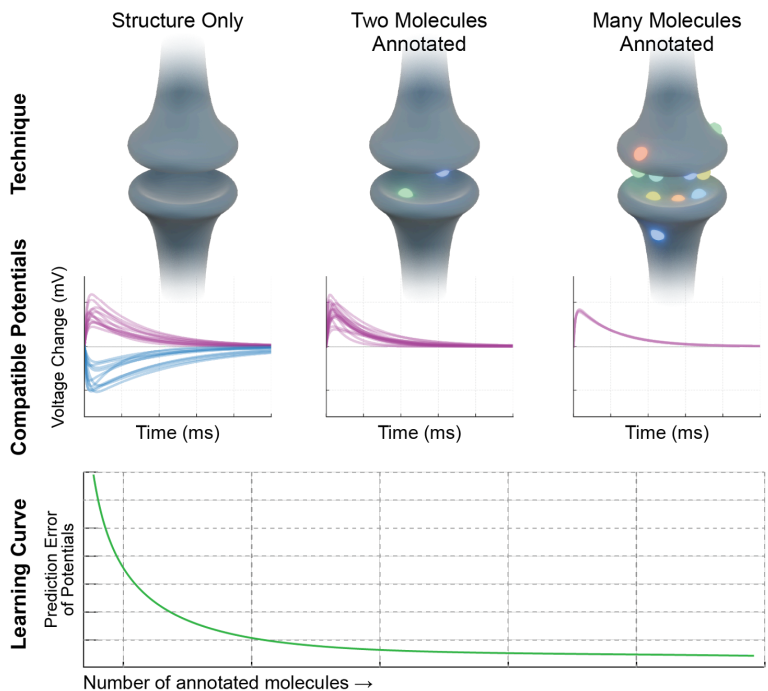

**Figure 1: Neural components can be imaged with high-dimensional molecular content and nanometer-scale ultrastructure, which can constrain predictions of postsynaptic potentials.** (Left) With EM, we can see synapses and their sizes, shapes, and connectivity. Many synaptic responses may be compatible with a given volume reconstruction, including active versus silent. (Middle) With expansion microscopy, we can quantify molecular content and aspects of molecular organization, including nanoscale spatial relationships among synaptic components; fewer physiological regimes will be compatible with such an image. (Right) The Ultrastructure-to-dynamics compiler converts molecular and ultrastructural measurements into model parameters (for example, synaptic conductances and kinetics) and their stochasticity (not shown), enabling forward prediction of physiological responses.

# Why now: imaging throughput and cost scaling

The compiler idea has been in the air for years[28], but it becomes actionable only when molecularly annotated ultrastructure can be collected cheaply enough to serve as key data, and when perturbation-labeled physiology can be gathered at sufficient scale for supervision.

The relevant scaling trends[29] between physiology and imaging are diverging. Neural recordings continue to improve, roughly doubling in scale every few years[30] (see ongoing tracking: https://stevenson.lab.uconn.edu/scaling/), but in vivo measurements face hard physical and practical constraints, including scattering and absorption, heating and phototoxicity limits, restricted fields of view, and challenges of wiring, stability, and spike sorting [31]. By contrast, fixed tissue can be partitioned and processed in parallel, and molecular imaging is rapidly gaining throughput and multiplexing capacity (see figure 2). Structural data will not replace dynamical measurement, but it can reduce ambiguity when paired with physiological ground truth. The practical implication is that imaging can plausibly supply the input side of a supervised compiler at the scale needed for learning, while perturbation-rich physiology supplies the labels.

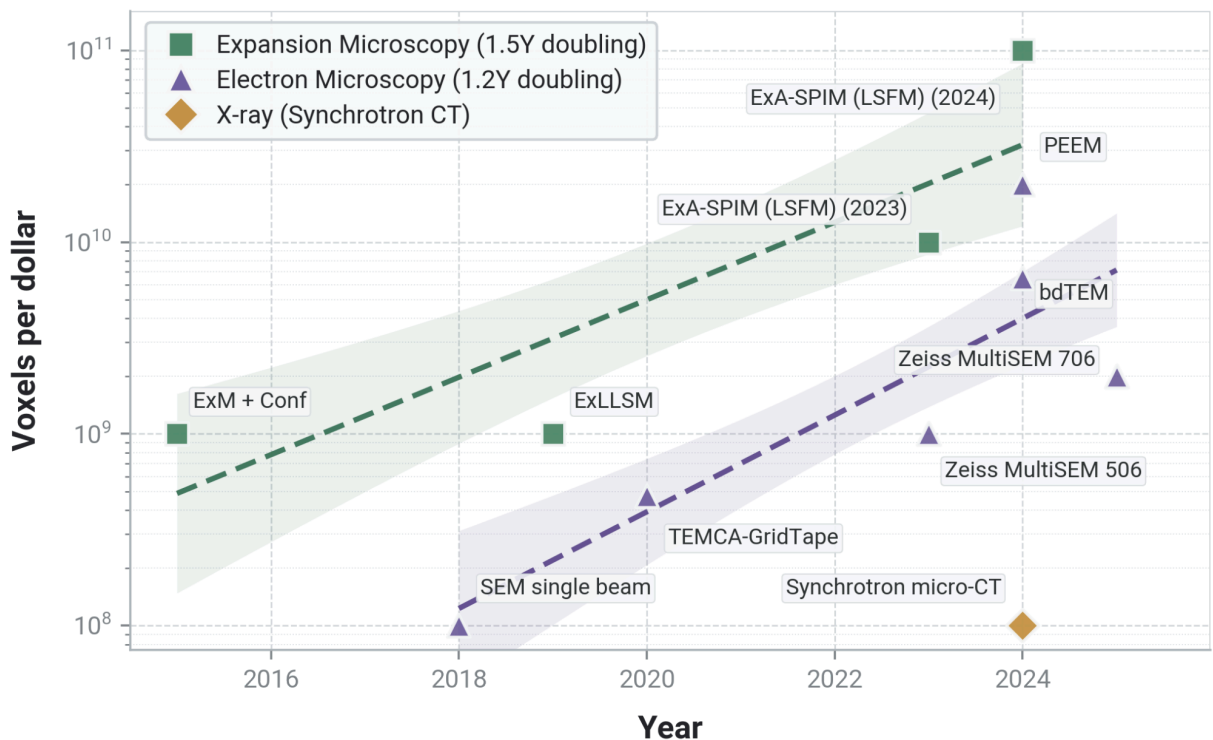

**Figure 2: Rapid exponential growth in anatomical imaging efficiency.** Note that the type of data is different across many of these studies, e.g., expansion with potential for multiple molecules imaged or EM with very high

resolution. For scale, the amount of data from electrical recordings doubles about every seven years, less than a factor of four over the shown interval[30].

At first glance, this creates a paradox for supervised learning: if training requires paired structure and dynamics, does slow physiology cap progress? The resolution is that training a compiler is not the same as executing it. We do not need dynamics for every circuit we wish to model; we need the dynamics of the components, such as synapses and neurites. High-throughput perturbation-and-readout paradigms (automated optical electrophysiology, multiplexed reporters) are beginning to be able to produce such corpora. Once trained, the compiler can be applied wherever structural maps are available, leveraging the faster scaling of imaging to extend mechanistic prediction to regimes where direct recording remains impractical. But this scaling argument only holds if the paired corpora needed for training actually exist — and currently they do not.

## The ground-truth bottleneck

We lack standardized datasets that co-register molecularly annotated ultrastructure with local dynamical measurements under controlled perturbations. This matters because the goal is identifiability of effective parameters[32]. Today, we usually cannot infer those parameters from structure alone, for two reasons. First, connectomic wiring without molecular readouts is an insufficient source language for the relevant degrees of freedom; for example, FlyWire[33] provides extraordinary anatomy and connectivity, but without molecular annotation, it omits key determinants of spiking, integration, and state dependence. Second, we lack paired training corpora that link richer structural measurements to perturbation-labeled dynamical ground truth at scale. Without both, the ultrastructure-to-parameter translation cannot be learned and benchmarked.

In the absence of such paired data, some recent work fills the gap with an alternative kind of supervision. Lappalainen et al.[34] illustrate a workaround: fix a computational objective, measure anatomy, and optimize missing parameters so the model performs the task. This can infer effective parameters, but generalization is limited by the assumed objective; change the task, and the parameters may need re-fitting. An ultrastructure-to-dynamics compiler aims to infer parameters from molecularly grounded measurements instead, reducing reliance on task assumptions and improving transfer across circuits and perturbations.
The goal is not to reconstruct every microscopic mechanism. The inferred parameters define an equivalence class: they summarize the net effect of unresolved details and are judged by predictive validity, not by one-to-one correspondence with a specific biophysical microquantity.

The question is therefore not whether perfect ground truth exists, but whether sufficient paired data can be assembled to make the translation learnable.

## From images to dynamics

Granting that the bottleneck is data, the question becomes: what evidence supports the possibility of learning, the possibility of compiling? AlphaFold[35] offers a useful precedent: the decisive ingredient was a large standardized corpus paired with benchmarks that rewarded generalization. Here too, progress will be limited less by model class than by paired corpora: co-registered structural inputs and perturbation-rich dynamical labels. Several efforts already align structure and dynamics at scale (for example, MICrONS[36]), but most current datasets lack either the molecular specificity, the biophysical targets (for example, synaptic and membrane currents), or the intervention-rich ground truth needed to evaluate parameter inference. This sets up two complementary paths, end-to-end learning and feature-based mechanistic inference (see Fig. 3).

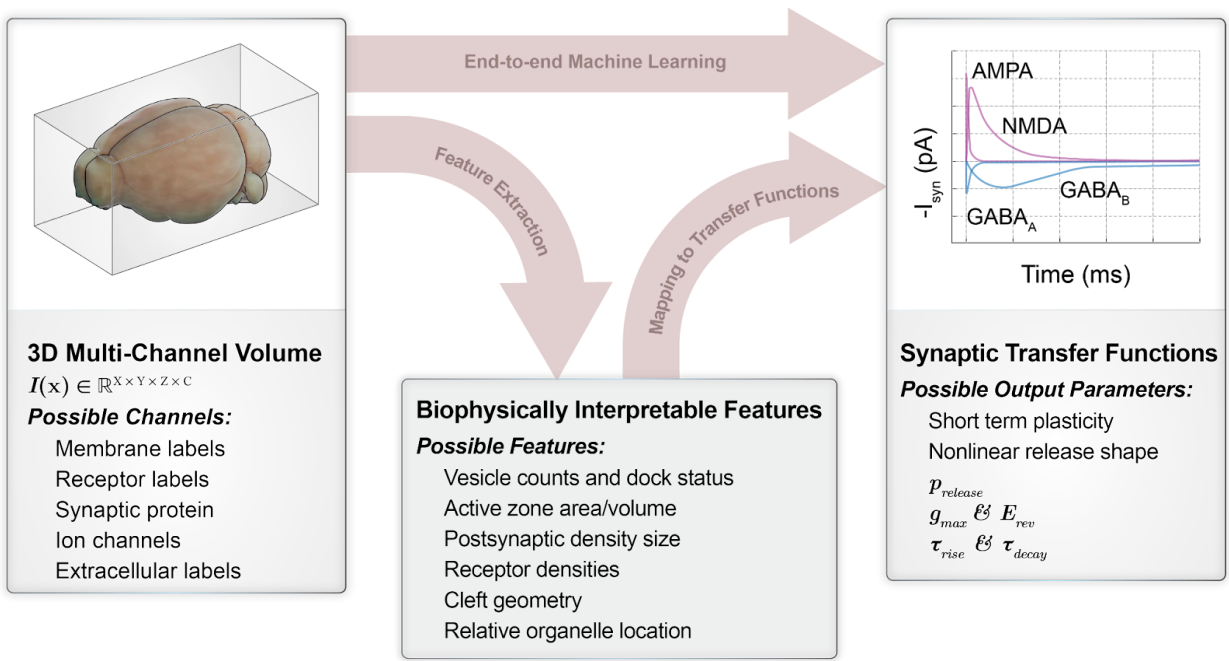

**Figure 3: Obtaining biophysical parameters from imaging.** We start with the 3D volume, containing labels for membranes, receptors, synaptic proteins and other markers. There are two paths towards calibrating parameter inference. One is an end-to-end machine learning approach where we simply train with enough data. There is an alternative path where we extract biologically meaningful features, build these into biologically interpretable models, and obtain biophysical parameters by modeling. A compiler would need to choose a timescale: discrete, continuous, or even just feedforward. Either way, we can estimate biophysical parameters based on the imaged structure.

Several adjacent results suggest the possibility of key components of this approach, even if no current dataset supports end-to-end compilation. Ultrastructural and molecular features carry functional signals[4,5,21], and when physiology anchors parameters, mechanistic models can be fit at scale[7,11] and informed by transcriptomic priors[12]. Emerging hybrid workflows pair recordings with post hoc molecular/ultrastructural readout in the same cells[37–40].

## The training data generation process

A compiler needs supervised examples: the same piece of tissue must be (i) imaged with molecularly annotated ultrastructure and (ii) assayed for local dynamics under controlled interventions (e.g., optophysiology, patch-clamping, etc). The scale required is not arbitrary — useful generalization likely requires paired measurements across hundreds of synapse classes and states, and cell types and states, spanning multiple perturbation conditions per class. Current automated optical electrophysiology and multiplexed reporter platforms[41,42] are approaching the throughput needed to produce corpora of this size, making the data generation problem demanding but not intractable.

**Perturb.** Perturbations should be chosen to isolate families of parameters while preserving biological context. Optical control of membrane voltage provides temporally precise probing of excitability and synaptic dynamics[41], and optogenetic or chemogenetic control of specific signaling pathways can interrogate protein-level mechanisms in situ[42]. Pharmacological and genetic perturbations provide complementary, often slower but more stable shifts in channel and receptor function. The aim is not exhaustive perturbation of every circuit of interest, but a transferable library of perturbation protocols that expose the dependence of model parameters on molecular and ultrastructural features.

**Measure.** Biophysical readouts must keep pace with perturbation throughput, motivating a shift from predominantly manual electrophysiology toward automated and optical measurements. Voltage imaging and targeted electrophysiology can provide synapse- and compartment-level constraints on conductances, kinetics, and short-term plasticity; calcium imaging can supply lower-bandwidth constraints when higher-resolution readouts are infeasible. Scalable perturbation-plus-readout paradigms, such as Perturb-seq[43], illustrate how systematic perturbation can be coupled to high-throughput measurement. Increasing multiplexed reporters further allows simultaneous measurement of multiple signals, enabling richer causal constraints on how molecular state shapes dynamics[44–46]. Recent advances in voltage imaging add access to spike timing and subthreshold dynamics that are central to mechanistic parameter inference[47–49].

**Building the Compiler using machine learning.** We will train a conditional generative translation from molecular ultrastructure to model parameters and stochastic components, with calibrated uncertainty. The core ML challenges — distribution shift across preparations and species, identifiability under partial observation, and the choice of whether to learn end-to-end or through biophysically interpretable intermediate features (Fig. 3) — are real but secondary to the data bottleneck: with sufficiently rich paired corpora, model class has historically mattered less than benchmark quality. Benchmark datasets, standardized alignment, and QC pipelines are therefore parts of the deliverable, not an implementation detail.

The compiler needs a first falsifiable test. The retina is a strong candidate: photons in, spike trains out, well-characterized cell types, and dense molecular annotation now within reach[36,50]. If ultrastructure cannot be compiled into models even there, the program fails. *C. elegans*[51] offers another test at whole-organism scale, probing whether molecular annotation improves prediction beyond what a known connectome alone provides. Organoids or cortical slices offer further simplified and smaller possibilities[52]. The key is to start where structure, dynamics, and behavior are most tractable, and where failure would be most informative. Biology has repeatedly advanced by modeling aspects of simple systems deeply. A comparable success for neural dynamics would be catalytic.

# Conclusion

The ultrastructure-to-dynamics compiler should be a major goal for the field: turn molecularly annotated ultrastructure into simulator-ready distributions over conductance-equivalent parameters, and judge success by out-of-distribution prediction, especially under held-out perturbations. If it works (see appendix for an extensive discussion of potential problems), static maps become executable models that predict responses, plasticity, and intervention effects. If it fails, it will likely reveal a wealth of information about which observables or abstractions are missing.

The bottleneck is shared calibration data, not ideas. Progress requires paired corpora that co-register structure with perturbation-labeled local dynamics, plus benchmarks and uncertainty-calibration norms. Those are infrastructure deliverables that do not fit the incentives of single-lab projects. Leadership therefore means building and maintaining this compiler stack as a community resource through a purpose-built, long-horizon effort (institute, consortium, focused research organization, or mission-driven startup). The maps are arriving; we need to build the infrastructure to make them speak.

# Acknowledgments

We are thankful to the many colleagues who provided feedback about this text.

# Imaging Technology Throughput Over Time

**Table A1: Imaging cost data.** Rough data on the various published imaging technologies. Many of these numbers come from a crowdsourcing effort: we thank the many involved scientists for providing estimates of these numbers. No claim beyond rough size (order of magnitude) is made.

| Year | Modality | Device Price ($K) | Total Cost ($/h) | Throughput (vox/h) | Approx. Voxels / $ | Contrast Source | Source |
|---|---|---|---|---|---|---|---|
| 2015 | ExM + Confocal | 500 | $125 | 10^11 | 10^9 | Any dye/antibody | 53 |
| 2024 | ExA-SPIM (LSFM) | 500 | $125 | 10^13 | 10^11 | Any dye/antibody | 54 |
| 2018 | EM (SEM, single beam) | 500 | $100 | 10^10 | 10^8 | Heavy Metal | 55 |
| 2019 | ExLLSM | 500 | $125 | 10^11 | 10^9 | Any dye/antibody | 56 |
| 2023 | ExA-SPIM (LSFM) | 500 | $125 | 10^12 | 10^10 | Any dye/antibody | 57 |
| 2023 | EM (Zeiss, MultiSEM 506) | 7000 | $575 | 10^11 | 10^9 | Heavy Metal | 58 |
| 2025 | EM (Zeiss, MultiSEM 706) | 10000 | $575 | 10^11 | 10^9 | Heavy Metal | 58 |
| 2024 | Synchrotron X-ray micro-CT | 2000 | $175 | 10^10 | 10^8 | Heavy Metal | 59–61 |
| 2024 | Benchtop mesoSPIM | 200 | $95 | 10^11 | 10^9 | Any dye/antibody | 62 |
| 2024 | PEEM | 2000 | $275 | 7x10^12 | 2x10^10 | Heavy Metal | 63 |
| 2024 | bdTEM | 500 | 125 | 8x10^11 | 6.5x10^9 | Heavy Metal | 64 |

# Appendix 1: potential limitations

## Scientific and conceptual limits

*Underdetermination persists even with "complete" structure.* Many different mechanistic parameterizations can produce similar dynamics. Molecular inventories may still leave out functional degrees of freedom (e.g., state dependence, phosphorylation, subunit composition, neuromodulatory tone). It is likely that some of these are not detectable with the current generation of antibodies or may even be lost post-mortem.

*Static snapshots may not predict dynamic state.* Imaging captures fixed tissue. But excitability and synaptic efficacy depend on recent activity, ongoing neuromodulation, metabolic state, and plasticity. Plasticity depends on history, spike timing, neuromodulators, and local protein state, not just static inventories. It is unknown how much these elements are reflected in structure.

*Molecule presence is not molecule function.* Protein counts do not guarantee effective conductance or kinetics. Trafficking, anchoring, subunit assembly, and local regulation can dominate. "Receptor number" is not the same as "synaptic weight." Machine learning may still extract predictive signal from these measurements, but only if the observed variables constrain effective parameters with sufficient precision. Moreover, a realistic resolution would not show the exact spatial locations of each molecule which may matter. This is one of the unknowns here - what resolution and what molecule diversity is needed for good performance.

*Choosing the right level of abstraction is tricky.* Too detailed and you cannot power the machine learning or construct feasible models; too abstract and structure stops constraining function.

*"Structure constrains function" may be true only in narrow regimes.* It may work well for some synapse classes and cell types (e.g., stereotyped circuits), but in the heterogeneous mammalian cortex, where compensation and variability are large, it may not be practically tractable to collect sufficient molecular and structural data for structure to be adequately constraining.

## Data and measurement limits

*Calibration data may be the true bottleneck.* Here we lean heavily on "paired structure-function" ground truth, but collecting it at necessary breadth, throughput, and standardization may be far harder than anticipated. Are we predicting currents or voltages? Under which circumstances? While what is being held constant? How do we then generalize?

*Measurement-induced artifacts and distortions.* Fixation, expansion, labeling, antibody access, clearing, and EM segmentation errors can systematically bias inferred parameters. *Generalization risk across brain regions, species, and conditions.* A translation learned in one animal, circuit, developmental stage, or preparation may fail elsewhere. Domain shift could be severe (cell types, temperature, concentrations of ions, signaling molecules and metabolites, neuromodulators, myelination, disease states). Although many molecular functions are conserved across species, it remains unclear whether a compiler trained in one species would transfer well to another.

## Modeling and scaling limits

*Compounding errors.* The compiler will inevitably exhibit estimation errors in local parameters such as currents and conductances. We don't know how such errors would compound. We know that in neural networks compounding is not catastrophic (and we can e.g., transfer a neural network to lower bit-depth[65]) but it is unknown if this would generalize to brains. At a minimum we should expect that we want to calibrate with structure-to-function and then re-calibrate[66] with some more large scale measurements (e.g., average activities).

*Model identifiability and validation could be ill-posed.* If models are flexible enough, they can fit calibration data but still fail out-of-distribution. "Predictive" could collapse into "interpolative," especially if benchmarks are not adversarial.

*Debuggability without mechanistic understanding may be a bottleneck.* Even if we can map ultrastructure and many molecular inventories, important functional state can live in slow biochemical history and regulatory context (hours-scale plasticity windows, transcriptional state, glial rhythms), and in highly context-specific signaling cascades. If a model misses a target behavior or perturbation response, it may be unclear whether the problem is missing measurements, missing model classes, or simply insufficient fidelity. The practical risk is that "more structure" does not monotonically translate into "more predictive," unless paired ground truth and perturbation-rich benchmarks make the translation identifiable and falsifiable.

*Computational cost could dominate the program.* Even if parameters are known, modeling large circuits with realistic dynamics may be prohibitive. There's a risk of shifting from an inference bottleneck to a compute bottleneck (but see [24]).

*Wireless and non-synaptic signaling can be difficult.* Diffusion, volume transmission, astrocytes, vascular coupling, immune effects, and extracellular space properties may be essential, but it is unknown how many of these effects would be reflected in molecular imaging.

## Translational and field-level limits

*Limits for clinical leverage.* Predicting drug response from tissue-level models requires bridging to pharmacokinetics, cell-type specificity, network compensation, and behavior.

*Ambiguity about success criteria.* Without explicit target metrics (e.g., predict PSP amplitude within X%, predict firing rate under Y perturbations, predict connectivity weights), the program risks becoming unfalsifiable. There is active work trying to improve such evaluations (see [67]). Success should imply the ability to predict behavior and activities but also to predict the effects of arbitrary system perturbations (lesions, optogenetics, etc.). Recent high-profile announcements in adjacent areas[68,69] have illustrated this risk concretely: without pre-specified, independently verifiable metrics, even technically interesting results attract legitimate skepticism that is difficult to rebut.

*Incentive failure mode.*
 The vision of structure-to-function demands large, standardized, infrastructure-heavy calibration datasets. If incentives do not reward this work, the community may produce gorgeous maps, but no shared ground truth, and FROs may be one vehicle to improve this[70]. Arguably, current systems do not have the right incentives.

*There are significant ethical and regulatory questions.* The idea of modeling brains clearly has ethical issues - if we simulate a being that we ascribe feelings to (say a cat), does it deserve the same protections as a real cat? There may also be regulatory issues, like who owns the content of brains? The ethical questions require in-depth analysis as we get closer to working simulations.

*Relation to the Blue Brain Project.* Ultrastructure-to-dynamics might appear to resemble earlier bottom-up modeling efforts, including the Blue Brain Project (BBP), in its ambition to connect biology to computation through mechanistic models[71]. The critical difference is what is treated as the primary scientific object. Much of BBP's visible output was modeling infrastructure, while the key missing ingredient, a shared, perturbation-anchored parameter corpus that makes models identifiable and comparable, remained scarce. Our proposal inverts the order. The centerpiece is a community calibration stack: co-registered molecular ultrastructure plus local dynamical labels under standardized perturbation suites, paired with benchmarks that require out-of-distribution generalization and uncertainty calibration. In this framing, models are interchangeable backends, and the field's progress is measurable early by whether inferred model parameters transfer across perturbations, preparations, and labs.

*Can we answer a specific question faster without compilers?* Perhaps. For some questions, direct experimental or model-based approaches may be faster. The value of a compiler is different: it is an enabling technology that makes many downstream questions answerable from structure in a reusable way.

# Appendix 2: Some potential uses of compilers

**Single cell modeling: e.g. how can we simulate a neuron to predict possible behaviors?**

Imagine you are a scientist specializing in modeling single neurons, say a ganglion cell in retina. To build these models[7,16] you usually assume that all channels are evenly distributed across the cell (usually not true), you assume that there are only two types of synapses from bipolar and amacrine cells, excitatory and inhibitory, and you assume activation kinetics (potentially from another species). You then simulate the neuron, aiming to elucidate how the neuron integrates information from its many synapses into neural spikes, but you know full well that the quality of the simulation depends on the precision of your estimates of synapses and other local dynamics.

Now, imagine the field had scanned the retina with many molecules, done the calibration work and machine learning and given you a compiler. You would choose one of the neurons in the retina (mind you that retinal neurons are not exact copies of one another). The compiler would give you the properties of the synapses that form the inputs to the neuron. It would also directly give you the nonlinear properties of the dendrites and axons. It would give you the kinetics. In other words it would give you all the parameters that your simulation needs.

Using the compiler output would move the model from a generic, physics-inspired simulation to a neuron-specific simulation. The results could then be tested in experiments (to be run before the reconstruction to be clear). The model simulation can thus be directly tested, it would make theories of information integration within neurons much more biologically meaningful.

**Microcircuits: how do local groups of neurons collaborate?**

Imagine you want to model a local microcircuit. You may know the cell types, morphology, and much of the connectivity. But you still lack the parameters that determine circuit dynamics: the strengths and kinetics of connections, dendritic nonlinearities, and how these vary across synapses, cells, and states. So even detailed microcircuit models still rely on sparse measurements and strong priors.

A compiler would estimate those missing local parameters directly from molecularly annotated ultrastructure. You could then simulate that specific circuit rather than an averaged or heavily fitted version of it. This would let you test whether recurrent amplification, normalization, excitation-inhibition balance, or other proposed motifs actually follow from the measured local biology[72]. The key advance is that microcircuit theories could be tested by prediction under held-out perturbations, not just by fitting observed activity.

**Opening new frontiers.**
Compilers would also improve existing large-scale models. In epilepsy, patient-specific models already combine anatomy and recordings, but local dynamical parameters are still usually imposed from generic priors. A compiler could replace those priors with parameters inferred from tissue-scale biology, improving forecasts of how resection, stimulation, or receptor-level interventions affect seizure propagation.

The same logic may matter for AI. Neural data has inspired architectures, but usually only at the level of broad motifs. A compiler could reveal local computational motifs and plasticity rules that are invisible in connectivity-only maps and erased in abstract neuron models. That may or may not help AI directly, but it would generate more biologically grounded hypotheses.

Whole-brain emulation is a longer-term possibility. In systems such as *C. elegans*, increasingly complete simulations already combine connectome, morphology, body, and environment[73]. A compiler would not solve the full problem, but it could solve one key part of it: estimating the local parameters needed to make structurally grounded simulations executable. And compilers have one further major advantage: physiology on humans is ethically problematic. But anatomy, the inputs to compilers, are much less problematic.

The point of a compiler is not to answer one question more quickly. It is to turn structural measurements into a reusable substrate for answering many mechanistic questions.